\documentclass{article}
\begin{document}
\begin{center}
\textbf{\large HIGGS BOSON MASS IN THE MINIMAL UNIFIED SUBQUARK MODEL}
\end{center}

\vspace{10mm}
\begin{center}
Hidezumi Terazawa 
\footnote{E-mail address: \textit{terazawa@mrj.biglobe.ne.jp}}
\end{center}

\vspace{5mm}
\begin{center}
\textit{Center of Asia and Oceania for Science(CAOS),}

\textit{3-11-26 Maesawa, Higashi-kurume, Tokyo 203-0032, Japan}

\textit{and}

\textit{Midlands Academy of Business \& Technology(MABT),}

\textit{Mansion House, 41 Guildhall Lane, Leicester LE1 5FR, United Kingdom}

\end{center}

\vspace{10mm}
\begin{center}
\textbf{\textit{Abstract}}
\end{center}

\vspace{5mm}
In the minimal unified subquark model of all fundamental particles and forces, the mass of the Higgs boson in the standard model 
of electroweak interactions($m_H$) is predicted to be about $2\sqrt{6}m_W/3$(where $m_W$ is the mass 
of the charged weak boson) so that $m_H = 131GeV$ for $m_W=80.4GeV$, to which the experimental values of 
$125-126GeV$ recently found by the ATLAS and CMS Colaborations at the LHC are very close.

\vspace{10mm}
What most of us can expect to find in high energy experiments at the Large Hadron Collider is the Higgs boson($H$), 
which is the only fundamental particle that has not yet been found in the standard model of electroweak interactions[1]. 
In the unified composite model of all fundamental particles and forces[2], the mass of the Higgs boson has been predicted 
in the following three ways: 

In general, in composite models of the Nambu-Jona-Lasinio type[3], the Higgs boson appears as a composite state of fermion-
antifermion pairs with the mass twice as much as the fermion mass. The unified subquark model of the Nambu-Jona-Lasinio type
[4] has predicted the following two sum rules: \[
m_W=[3(m_{w_1}^2+m_{w_2}^2)/2]^{1/2}\] and \[
m_H=2[(m_{w_1}^4+m_{w_2}^4)/(m_{w_1}^2+m_{w_2}^2)]^{1/2}, \]
where $m_{w_1}$ and $m_{w_2}$ are the masses of the weak-iso-doublet spinor subquarks called \lq\lq wakems\rq\rq\ standing 
for weak and electromagnetic($w_i$ for $i=1,2$) while $m_W$ and $m_H$ are the masses of the charged weak boson($W$) and 
physical Higgs scalar in the standard model, respectively. By combining these sum rules, the following relation has been 
obtained if 
$m_{w_1}=m_{w_2}$: \[
m_w:m_W:m_H=1:\sqrt{3}:2. \]
From this relation, the wakem and Higgs boson masses have been predicted as \[
m_w=m_W/\sqrt{3}=46.4GeV \] and \[
m_H=2m_W/\sqrt{3}=92.8GeV \]
for $m_W=80.4GeV$[5].
On the other hand,if $m_{w_1}/m_{w_2}=0$ or $m_{w_2}/m{w_1}=0$, the other relation can be obtained: \[
m_w:m_W:m_H=1:\sqrt{3/2}:2. \]
From this relation, the non-vanishing wakem and Higgs boson masses can be predicted as \[
m_w=m_W/\sqrt{3/2}=65.6GeV \] and \[
m_H=2m_W/\sqrt{3/2}=131GeV \]
for $m_W=80.4GeV$[5].
More generally, from the two sum rules, the Higgs boson mass can be bounded as \[
92.8GeV=2m_W/\sqrt{3}\leq m_H\leq 2\sqrt{6}m_W/3=131GeV. \]

In the unified quark-lepton model of the Nambu-Jona-Lasinio type[4], the following two sum rules for $m_W$ and $m_H$ have 
been predicted: \[
m_W=(3<m_{q,l}^2>)^{1/2} \] and \[
m_H=2(\sum m_{q,l}^4/\sum m_{q,l}^2)^{1/2}, \]
where $m_{q,l}$\rq s are the quark and lepton masses and $<>$ denotes the average value for all the quarks and leptons. If 
there exist only three generations of quarks and leptons, these sum rules completely determine the top quark and Higgs 
boson masses[6] as \[
m_t\cong (2\sqrt{6}/3)m_W=131GeV \] and \[
m_H\cong 2m_t\cong (4\sqrt{6}/3)m_W=263GeV. \]

Furthermore, triplicity of hadrons, quarks, and subquarks[7] tells us that these sum rules can be further extended to the 
approximate sum rules of \[
m_W\cong (3<m_{B,l}^2>)^{1/2} \] and \[
m_H\cong 2(\sum m_{B,l}^4/\sum m_{B,l}^2)^{1/2}, \]
where $m_{B,l}$s are the \lq\lq canonical baryon\rq\rq\ and lepton masses and $<>$ denotes the average value for all the 
canonical baryons and leptons. The \lq\lq canonical baryon\rq\rq\ means either one of $p,n$ and other ground-state
 baryons of spin 1/2 and weak-isospin 1/2 consisting of a quark heavier than the $u$ and $d$ quarks and a scalar and 
isoscalar diquark made of $u$ and $d$ quarks. If there exist only three generations of quarks and leptons, these sum rules 
completely determine the masses of the canonical topped baryon, $T$, and the Higgs scalar as \[
m_T\cong 2m_W=161GeV \] and \[
m_H\cong 2m_T\cong 4m_W=322GeV. \]

Therefore, if the Higgs boson is found with the mass between 92.8GeV and 131GeV, it looks like a composite state of 
subquark-antisubquark pairs. If it is found heavier with $m_H$ around 263GeV or even 322GeV, it can be taken as a bound 
state of $t\overline{t}$ (\lq\lq topponium\rq\rq) or $T\overline{T}$ (\lq\lq topped-baryonium\rq\rq), respectively. If it 
is found with the mass lying between these typical masses, it may be taken as a mixture of subquark-antisubquark pairs 
and quark-antiquark pairs, \textit{etc.}.

Very recently, the ATLAS and CMS Collaboration experiments at the CERN Large Hadron Collider have almost excluded the two ranges 
for the Higgs boson mass: the one lower than 114GeV and the other between 141GeV and 476GeV[8,9], which disagrees with 
both the prediction in the unified quark-lepton model of the Nambu-Jona-Lasinio type[4] and that in the unified baryon-lepton 
model of the Nambu-Jona-Lasinio type[7]. Instead, the prediction in the unified subquark model[4]($92.8GeV\leq m_H\leq 131GeV$) 
shows a right ballpark on which the mass of the Higgs boson in the standard model should land. Moreover, the fact that 
the experimental values of $m_H = 125-126GeV$ recently found by the ATLAS and CMS Collaborations are very close to the predicted 
one of $m_H = 2\sqrt{6}m_W/3 = 131GeV$ seems to strongly suggest that either $m_{w_1}/m_{w_2}$ or $m_{w_2}/m_{w_1}$ vanishes. 
It seems to indicate that the Higgs boson is a composite of the isodoublet spinor subquark-antisubquark pairs well described 
by the minimal unified subquark model with either one of subquark masses vanishing. Let us hope that the future LHC experiments 
will tell us whether the minimal unified subquark model is a viable model of all fundamental particles and forces!

\vspace{5mm}
\textbf{\large Acknowledgements}

\vspace{5mm}
The author thanks Professor Yuichi Chikashige for very useful helps in correcting errors in the original manuscripts.

\end{document}